\documentclass[10pt]{article}

\usepackage{graphicx}
\usepackage{color}
\usepackage{subfig}
\usepackage[percent]{overpic}
\usepackage[numbers, square, comma, sort&compress]{natbib}

\begin{document}

\title{Comments on Phys. Rev. D89, 097101 (2014) \\
{\large ``Reevaluation of the parton distribution of strange quarks 
in the nucleon''} }


\author{M. Stolarski
\footnote{Present address: LIP-Lisboa, Av. Elias Garcia 14 - 1$^{o}$ 
1000-149 Lisboa, Portugal } }

\maketitle


\begin{abstract}

The HERMES collaboration in Phys. Rev. D89, 097101 (2014)
extracted information about the strange quark density in the
nucleon. One of the main results is an observation that
the shape of the extracted density is very different
from the shapes of the strange quark density from global
QCD fits and also from that of the light antiquarks.  

In this paper systematic studies on the HERMES published 
multiplicity of pion and kaon data are presented. 
It is shown that the conclusions 
concerning the strange quark distribution in the nucleon
reached in Phys. Rev. D89, 097101 (2014) are at the moment
premature.

\end{abstract}

\section{INTRODUCTION}

The information on the strange quark properties
in the nucleon is rather scarce.
The HERMES results \cite{her2} on semi inclusive deep inelastic scattering
should bring a very important contribution
to this problem.
The collaboration recently confirmed a previous claim \cite{her1} 
that the strange quark parton distribution function (PDF) has very 
different shape in the Bjorken $x$ scaling variable as compared to the 
distribution of non-strange sea. This statement has deep impact in the field. 

Unfortunately both HERMES analyses \cite{her2} and \cite{her1} 
are based solely on the study of the kaon multiplicity sum, 
without presenting any further systematic checks. 
In this paper, after a brief explanation of the HERMES analysis method,
the pion multiplicity sum is discussed. 
As a self consistency check for the results obtained using 
the kaon multiplicity sum,
a different combination of $K^+$ and $K^-$ multiplicity is also studied.
Finally, the results concerning the  kaon multiplicity difference are presented.

\section{THE HERMES METHOD}

HERMES studied the semi-inclusive deep inelastic
scattering (SIDIS) of electrons impinging on a deuteron target.
The results \cite{her2} on the strange quark PDF $S(x)$ are based on the
analysis of the sum of multiplicities of charged kaons $(K^+ + K^-)$ as
a function of  Bjorken $x$, $dN^K(x)/dN^{DIS}(x)$.
As stated there, in LO pQCD,
\begin{equation}
\frac{dN^K(x,Q^2)}{dN^{DIS}(x,Q^2)} = \frac{Q(x,Q^2) \int D_Q^K(z,Q^2)dz 
+ S(x,Q^2) \int D_S^K (z,Q^2)dz}{5Q(x,Q^2) + 2S(x,Q^2) },
\label{eq:K}
\end{equation}
where $Q$ and $S$ are a combination of non-strange quark densities, 
$Q=u+\bar{u}+d +\bar{d}$ and of strange ones $S= s +\bar{s}$.
The $D_Q^K(z,Q^2)$ is a fragmentation function defined as 
$D_Q^K(z,Q^2) = 4D_u^K(z,Q^2) + D_d^K(z,Q^2)$, while
$D_S^K(z,Q^2) = 2D_s^K(z,Q^2)$. The $Q^2$ is the 
negative four momentum transfer and $z$ (in lab. frame) 
is the fraction of the photon energy carried by the hadron.

As presented in Fig.~1 of \cite{her2}, the kaon multiplicity is flat 
for high $x$, i.e., in the region where a small contribution from strange 
quarks is expected, and it rises by about 20-30\% for lower values of $x$.
Without strange quarks the distribution should be almost flat.
Therefore, the increase of the kaon multiplicity sum in the low $x$ region
is interpreted as a strong signature of the strange quarks presence
(see Figs. 2, 3 in Ref. ~\cite{her2}). In addition the $x$ dependence of 
the kaon multiplicity sum suggests
that the shapes of the strange and non-strange sea are very different.

\section{SYSTEMATIC STUDIES} 

\subsection{The $\pi$ multiplicity sum}

The multiplicity analyses are quite complex as they depend a lot on
various correction factors. For this reason
before claiming certain features of the strange quark distribution
using kaon multiplicities, as a natural self consistency check 
the pion results should be verified.
The multiplicity sum of pions can be written in a similar way as in 
Eq.(\ref{eq:K}),
\begin{equation}
\frac{dN^{\pi}}{dN^{DIS}}  =  \frac{Q\, D_Q^{\pi} + S\,  D_S^{\pi}} { 5Q + 2S},
\label{eq:pi}
\end{equation}
where $D_Q^{\pi}$ and $D_S^{\pi}$ are the pion fragmentation functions
defined in a similar way as for kaons. Here for simplicity the $x$, $Q^2$, $z$ 
dependence was
omitted and $D_i^{\pi} = \int D_i^{\pi}(z,Q^2) dz$. 
The Eq.~(\ref{eq:pi}) can be re-written in the following form:
\begin{equation}
\frac{dN^{\pi}}{dN^{DIS}} = \frac{D_Q^{\pi}}{5}  + \frac{ S }{ 5Q+2S } 
( D_S^{\pi} -0.4 D_Q^{\pi} ).
\end{equation}
It is interesting to notice that:  
\begin{equation}
 D_S^{\pi} -0.4 D_Q^{\pi}  = 2D_s^{\pi} -1.6 D_u^{\pi} - 0.4 D_d^{\pi} < 0.
 \label{eq:pi2}
\end{equation}
The right hand side of Eq.(\ref{eq:pi2}) is expected to be {\em negative} 
since $D_u^{\pi}$ and  $D_d^{\pi}$ involve
the so called favored fragmentation functions
$D_u^{\pi^+}$ and $D_d^{\pi^-}$.
{\color{black}In LO, due to the presence of strange quarks,
the pion multiplicity sum should decrease for lower values of $x$,
contrary to the kaon case. }

\begin{figure}[ht]
  \includegraphics[width=1.0\textwidth]{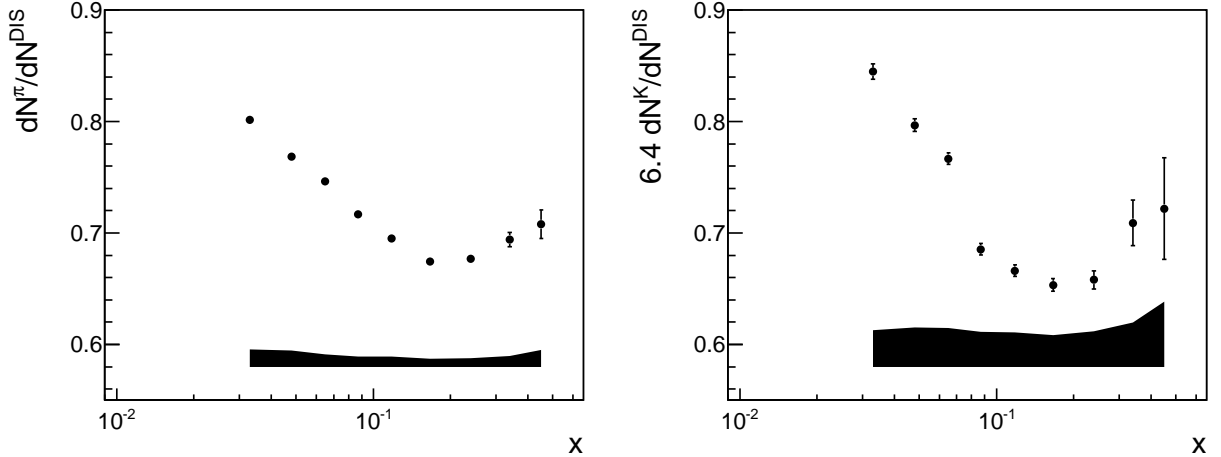}
  \caption{Comparison of $\pi$ and K multiplicity sums 
    obtained from \cite{her3}, left and right panels respectively. 
    Striking and not expected similarities in shape are observed.}
  \label{fig:1}
\end{figure}

The $\pi$ multiplicity sum can be extracted from the published HERMES data 
\cite{her3}.
The result is shown on the left panel of Fig.~\ref{fig:1},
using $x$ representation of the data (vector meson subtracted), 
{\color{black}the presented systematic uncertainty band takes into account
the correlation factors between systematic uncertainties of the 
$\pi^{+}$, $\pi^{-}$ data. These correlation factors were obtained indirectly
from the published uncertainties of $M^{\pi^{+}}$, $M^{\pi^{-}}$
and $(M^{\pi^{+}}- M^{\pi^{-}})/(M^{\pi^{+}}- M^{\pi^{-}})$~.
Correlation factors were also extracted for the kaon case, where they reach a 
value up to 0.85. }
Contrary to the expectations ($cf.$ above) the shape of the distribution 
is very similar to
the one of the $K$ multiplicity sum, presented in the right panel of 
Fig.~\ref{fig:1}. {\color{black}For easier comparison the kaon multiplicity
sum was multiplied by a factor 6.4, which minimizes the $\chi^2$ between
$\pi$ and $K$ multiplicity sums: a value of $\chi^2/ndf = 5.1/8$
in the case when the correlation between positive and negative particles is taken
into account as described above, or $\chi^2/ndf = 7.7/8$ in the case when systematic 
uncertainties are treated as independent.}
Especially interesting is the fact that
both multiplicities start
to rise more or less at the same $x$. In \cite{her2}, the increase for kaons is
attributed solely to the strange quark PDF having very different shape from
the non-strange sea. But as presented here, such an explanation does not work
for the $\pi$ data. On the other hand, very similar shape 
of the distributions suggests 
that there is a common source which causes the observed effect.

The $Q^2$ dependence of $D_Q^{\pi}$
in both LO and NLO is expected to be rather weak in the region of interest.
Thus it cannot explain the features observed in the left panel 
of Fig \ref{fig:1}. In the DSS fit of fragmentation functions \cite{dss}, 
both in LO and NLO  
$D_Q^{\pi}(Q^2=1 { \rm GeV^2}) < D_Q^{\pi}(Q^2=3 {\rm GeV^2})$.
The lowest HERMES $x$ corresponds to $\langle Q^2 \rangle \approx 1.2$ GeV$^2$,
while $\langle Q^2 \rangle=3$ GeV$^2$ for $x \approx 0.15$.
The weak $Q^2$ dependence is also seen in HERMES data in the $Q^2$ 
representation. 
The pion multiplicity sum is basically flat in the full $Q^2$ range with 
an average $dN^{\pi}/dN^{DIS}=0.717$.
So the effect presented in the left panel of Fig.~\ref{fig:1} 
is indeed related to $x$ or $y$ dependence, and not $Q^2$.

It is worth to mention that in Ref. \cite{lss01} problems with the perturbative
description of HERMES pion data in LO and NLO are discussed. 
Moreover, as noted by the same authors
in \cite{lss02}, the analyses of HERMES multiplicity data available 
in $x$ and $Q^2$ representations do not give compatible results.
For example for $x$ representation at
$x=0.17$ and $Q^2=3.2$ GeV$^2$ the $\pi$ multiplicity sum is:
$0.675 \pm 0.002 \pm 0.006$. For almost the 
same kinematic point, $x= 0.17$ and $Q^2=3.8$ GeV$^2$, but in $Q^2$ representation,
the $\pi$ multiplicity sum is $0.714 \pm 0.002 \pm 0.007$.
A difference of about 6\% is seen.
It is much larger than statistical and systematic uncertainties,
while a large part of the data is the same.
Observe that the average $x$ value of the presented data points is high enough, 
so that the point in $Q^2$ 
representation doesn't contain low $x$ data $(x<0.08)$, 
whose presence could otherwise explain the observed difference. 
In my opinion the HERMES multiplicities should be available
simultaneously in $(x,y,z)$ or $(x,Q^2,z)$ representations
so that such effects can be studied in more details.
In what follows it is assumed that the current HERMES results
in $x$ representation were extracted correctly from the data.

\subsection{Kaon multiplicities}

The analysis of the strange quark content of the nucleon
using the kaon multiplicity sum is one of many possible choices.
For a self consistency check one can use
other combinations of $K^+$ and $K^-$ multiplicities.
The drawback is that to have a simple formula one has
to assume in addition that $s(x)=\bar{s}(x)$
\footnote{It was verified that not using $s=\bar{s}$ assumption, 
but strange quark densities as in MSTW08 L0 PDF \cite{mstw}, the 
conclusions presented later are not changed.}.
One such possible choice is to cancel the contribution from 
$D_u^{K^{+}}$ in the expression for the multiplicity.
The starting point is the kaon multiplicity difference 
(see Ref. \cite{chlead}),
\begin{equation}
\frac{ dN^{K^{diff}} }{dN^{DIS}}= 
\frac{(u_v +d_v) ( 4 D^{K^+}_u - 4 D^{K^+}_{\bar{u}} + 
       D^{K^+}_d - D^{K^+}_{\bar{d}}) }{ 5Q+2S}.
\label{eq:Kdiff}
\end{equation}
To have a common notation, Eq.(\ref{eq:K}) can be rewritten as
\begin{equation}
\frac{ dN^K }{ dN^{DIS} }= \frac{Q( 4 D^{K^+}_u + 4 D^{K^+}_{\bar{u}} + 
       D^{K^+}_d + D^{K^+}_{\bar{d}})+ SD_S^K}{5Q+2S}.
\label{eq:Ksum2}
\end{equation}
Combining Eq.(\ref{eq:Kdiff}) and Eq.(\ref{eq:Ksum2}) one obtains
\begin{equation}
\frac{5Q+2S}{Q} \frac{dN^K}{dN^{DIS}} -  
\frac{5Q+2S}{u_v+d_v} \frac{dN^{K^{diff}}}{dN^{DIS}} 
=  \frac{dN^{K'}}{dN^{DIS}} =
8 D^{K^+}_{\bar{u}} + 2 D^{K^+}_{\bar{d}} + \frac{S}{Q} D_S^{K}.
\label{eq:Kprim}
\end{equation}

\begin{figure}
  \includegraphics[width=1.0\textwidth]{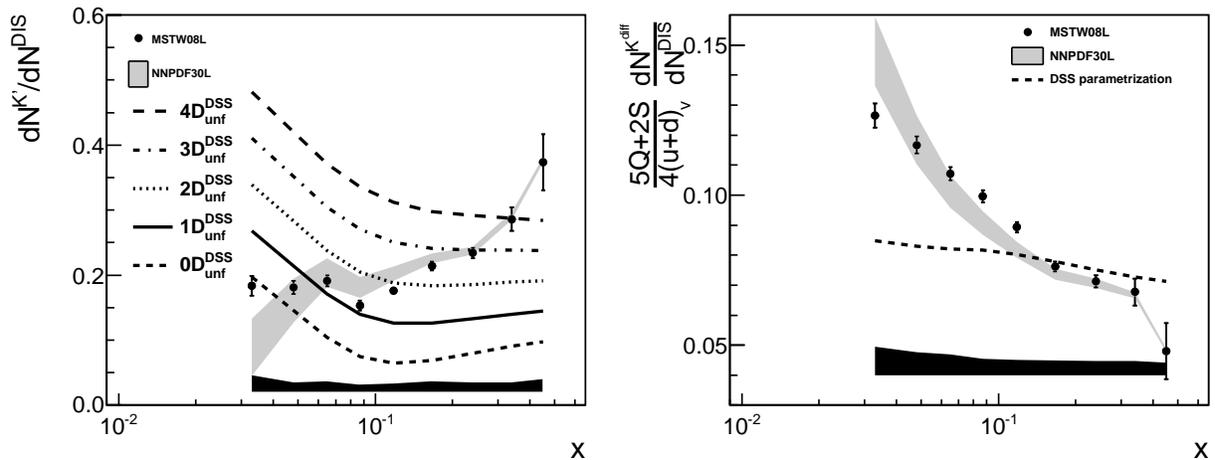}
  \caption{
    Left panel: A tentative of extraction of the strange quark 
    contribution in the nucleon using a certain combination of 
    $K^+$ and $K^-$ multiplicities $(K')$, where the contribution from 
    $D_{u}^{K^{+}}$ cancels. 
    A drop of the multiplicity at low $x$ is observed instead
    of the expected rise. The points correspond to the $K'$ extraction
    using MSTW08 LO, and black band on the bottom marks systematic 
    uncertainty of data. The $K'$ was also extracted using NNPDF30L set,
    the gray band it the  $\pm 1 \sigma$ uncertainty of NNPDF30L.
    Finally, a set of lines corresponds
    to the LO prediction for $K'$, where on top of extracted from HERMES
    $S(x) \int D_{str}$ distribution (divided by $Q$) a $10D^{DSS}_{unf}$ is added 
    with additional scaling factors applied at the reference scale.
    Right panel: Comparison of the kaon multiplicity 
    difference in the HERMES data multiplied by $(5Q+2S)/4(u_{v}+d_{v})$ 
    with the DSS expectation. }
  \label{fig:2}
\end{figure}

The idea of the analysis is exactly the same as for the kaon multiplicity sum. 
Namely an increase of the multiplicity should be observed for low $x$, due
to the increased strange quark presence.
Moreover, as in the original idea of the HERMES analysis in Ref. \cite{her2} 
one does not need to know separately 
$D^{K^+}_{\bar{u}}$ and $D^{K^+}_{\bar{d}}$, 
{\color{black}but only
their combination presented in Eq.(\ref{eq:Kprim}). 
This combination is extracted from the data at high $x$, 
exactly as was
done for $D_Q^K$ in \cite{her2}. }
Based on the results presented
in Fig.~3 of \cite{her2}, at lowest $x$ one expects a rise 
of  $dN^{K'}/dN^{DIS}$ due to $SD_S^{K}/Q$ by about 0.18$\,$~.
The actual results of $dN^{K'}/dN^{DIS}$ are presented in the left panel 
of Fig.~\ref{fig:2}. The MSTW08 LO PDF \cite{mstw} 
was used in the evaluation of results. 
Contrary to the expectations, the  multiplicity decreases. 
As it is hard to expect that the $Q^2$ dependence of 
$D^{K^+}_{\bar{u}}$ + $D^{K^+}_{\bar{d}}$ can fully explain the shape 
presented in Fig.~\ref{fig:2},
this is an indication of a failure of the conventional LO pQCD parton model
approach in the analysis of multiplicities at HERMES kinematics, 
see also Ref. \cite{aram}.
For completeness, the extraction of $dN^{K'}/dN^{DIS}$ 
using NNPDF30L \cite{nnpdf} was also added on Fig.~\ref{fig:2}.
The gray band marks $\pm 1\sigma$ of the NNPDF30L uncertainty.
Finally, a set of lines corresponds to the expected value
of $dN^{K'}/dN^{DIS}$ in LO, assuming $S(x) \int D_{str}$ 
as extracted by HERMES and a DSS parametrization
of $D_{unf}$, which at the reference scale was multiplied 
by a factor in the range $\in (0,4)$. To perform the $Q^2$ evolution
of the re-scaled FF a {\sc QCDNUM} program was used \cite{qcdnum}.

The problem presented in the left panel of Fig.~\ref{fig:2} 
is also seen while studying just the
kaon multiplicity difference. 
As can be deduced from Eq.(\ref{eq:Kdiff})
under the assumption $s(x)= \bar{s}(x)$, the kaon multiplicity difference 
does not depend upon the strange quarks density, 
but it is obviously correlated with $D_Q^K$. 
Multiplying the kaon multiplicity difference by $(5Q+2S)/4(u_v+d_v)$,
one effectively obtains what in the DSS fit can be related to the difference 
of the so called favored and unfavored FF 
(observe that in  DSS $\,D^{K^+}_d - D^{K^+}_{\bar{d}}=0$).
According to the DSS fit this combination
has a weak $Q^2$ dependence of  about 10\%. What is actually 
found in the HERMES data
using $x$ representation (vector meson subtracted)
is presented in the right panel of Fig.~\ref{fig:2}. 
There is a clear disagreement between the HERMES data and the 
{\color{black}weak $Q^2$ dependence expected from the DSS fit. }
For completeness the exercise was repeated with NNPDF30L set,
$\pm 1 \sigma$ of PDF uncertainty around extracted values 
is shown as a gray band on the right panel of Fig.~\ref{fig:2}.

The $dN^{K^{diff}}/dN^{DIS}$ is strongly correlated with $D_Q^K$,
{\color{black} i.e., in LO pQCD, assuming $s(x)= \bar{s}(x)$, 
$D_Q^K$ can be expressed as a function of $dN^{K^{diff}}/dN^{DIS} \cdot (5Q+2S)/(u_v+d_v)$
and some unfavored FF. }
One cannot assume that the reason causing the unexpected $x$ dependence
of $dN^{K^{diff}}/dN^{DIS}$ is not affecting the extraction of $D_Q^K$. 
In such a situation using Eq.(\ref{eq:K}) to
extract $S \, D_S^K$ as done in \cite{her2} is not justified. 
It is also clear that even if one tries to explain the observed
rise in the kaon multiplicity sum as solely related to the strange quarks, 
the observed features in the kaon multiplicity difference, $dN^{K'}/dN^{DIS}$,
and in the $\pi$ multiplicity sum will not be explained.

\section{SUMMARY} 

It is possible that the strange quark distribution in the nucleon
has indeed different shape than the non-strange sea.
However, the tests presented in this paper indicate that 
the analysis of HERMES
data \cite{her3} based on the conventional LO pQCD parton model approach 
cannot be used to support the final conclusion of their paper \cite{her2}.
Namely that the shape of  $xS(x,Q^2)$ is strikingly different
from that of global QCD fits and the sum of the light antiquarks. 
The above conclusion assumes that multiplicities published in \cite{her3} 
were extracted correctly from the data.
It would be clearly beneficial to the community if HERMES multiplicity 
data were available simultaneously in $(x,Q^2,z)$ or $(x,y,z)$ intervals
so that more systematic tests could be done in the future.


\section*{ACKNOWLEDGMENTS}
The author thanks A. Kotzinian for his comments and remarks concerning the manuscript.  
This research was supported by the Portuguese Funda\c{c}\~ao para a Ci\^encia 
e Tecnologia, grant SFRH/BPD/64853/2009.


\begin{thebibliography}{99}

{\small

\bibitem{her2} A. Airapetian {\em et al.} (HERMES Collaboration), 
Phys. Rev. {\bf D89}, 097101 (2014).


\bibitem{her1} A. Airapetian {\em et al.} (HERMES Collaboration), 
Phys Lett. {\bf B666}, 446 (2008).


\bibitem{her3} A. Airapetian {\em et al.} (HERMES Collaboration), 
Phys. Rev. {\bf D87}, 074029 (2013).

\bibitem{dss}  D. de Florian, R. Sassot and M. Stratmann,
Phys Rev {\bf D75}, 114010 (2007).  


\bibitem{lss01} E. Leader, A. V. Sidorov, and D. B. Stamenov, 
Phys. Rev. {\bf D90}, 054026 (2014).

\bibitem{lss02} E. Leader, A. V. Sidorov, and D. B. Stamenov,
{\it Proceedings of the XV
Advanced Research Workshop on High Energy Spin Physics}, (DSPIN-13), 
October 8-12, 2013,  Dubna, Russia, pp 124-130, 
[hep-ph 1312.5200].


\bibitem{mstw} A. Martin, W. Stirling, R. Thorne, and G. Watt, 
Eur. Phys. J. {\bf C63}, 189 (2009).


\bibitem{chlead} E. Christova and E. Leader, 
Phys. Rev. {\bf D79}, 014019 (2009).

\bibitem{aram} A. Kotzinian, Eur. Phys. J. {\bf C44}, 211 (2005).

\bibitem{nnpdf} R. D. Ball {\em et al.} (NNPDF Collaboration), hep-ph 1410.8849 
to be published in JHEP.


\bibitem{qcdnum} M. Botje, Comput. Phys. Commun. {\bf 182}, 490 (2012).



}

\end{thebibliography}
\end{document}